# Temperature dependence of the damping parameter in the ferrimagnet $Gd_3Fe_5O_{12}$

Isaac Ng,[1,2 a)] Ruizi Liu[1,3 a)], Zheyu Ren[1,3], Se Kwon Kim,[4] and Qiming Shao [1,2,3 b)]

[1]Department of Electronic and Computer Engineering Department, Hong Kong University of Science and Technology, Clear Water Bay, Kowloon, Hong Kong SAR, China

[2]Department of Physics, Hong Kong University of Science and Technology, Clear Water Bay, Kowloon, Hong Kong SAR, China

[3]Guangdong-Hong Kong-Macao Joint Laboratory for Intelligent Micro-Nano Optoelectronic Technology, The Hong Kong University of Science and Technology, Hong Kong, China

[4]Department of Physics, Korea Advanced Institute of Science and Technology, Daejeon 34141, Republic of Korea

[a)] Contributed equally [b)] Email: eeqshao@ust.hk

**Abstract**

The damping parameter $\alpha_{\text{FM}}$ in ferrimagnets defined according to the conventional practice for ferromagnets is known to be strongly temperature dependent and diverge at the angular momentum compensation temperature, where the net angular momentum vanishes. However, recent theoretical and experimental developments on ferrimagnetic metals suggest that the damping parameter can be defined in such a way, which we denote by $\alpha_{\text{FiM}}$, that it is free of the diverging anomaly at the angular momentum compensation point and is little dependent on temperature. To further understand the temperature dependence of the damping parameter in ferrimagnets, we analyze several data sets from literature for a ferrimagnetic insulator, gadolinium iron garnet, by using the two different definitions of the damping parameter. Using two methods to estimate the individual sublattice magnetizations, which yield results consistent with each other, we found that in all the used data sets, the damping parameter $\alpha_{\text{FiM}}$ does not increase at the angular compensation temperature and shows no anomaly whereas the conventionally defined $\alpha_{\text{FM}}$ is strongly dependent on the temperature.

Antiferromagnets have been one important focus in spintronics due to their properties distinct from more conventional ferromagnets including the zero stray field, ultrafast dynamics, and immunity to external field[1,2]. Recently, antiferromagnetically coupled ferrimagnets have emerged as a new material platform to study antiferromagnetic dynamics as suggested by the recent discoveries of current-driven magnetization switching near magnetization compensation point[3,4], where the net magnetization vanishes, and fast domain-wall dynamics at the angular momentum compensation temperature[4,5,6,7], where the net angular momentum vanishes. However, magnetic resonance and dynamics of ferrimagnets have not been fully understood partly due to the involvement of multiple magnetic sublattices and the resultant internal complexity. The dissipation rate of angular momentum in magnetic material is manifested as the linewidth in resonance spectrum. One quantity of particular importance in the dissipative dynamics of ferrimagnets is the damping parameter, which is a characteristic of the magnetic material that determines the Gilbert-like damping of angular momentum and is usually denoted by the dimensionless number $\alpha$. Early literature suggested that the effective damping parameter $\alpha_{FM}$ for ferrimagnets defined by a value that is proportional to the linewidth of the resonance response is strongly temperature-dependent and increases anomalously near the angular momentum compensation temperature ($T_A$).[8] Recent studies have provided a new interpretation: the damping parameter can be defined in such a way that it is independent of temperature near the $T_A$ while the temperature dependence of the ferromagnetic resonance (FMR) is attributed to the temperature dependence of the net angular momentum.[9,10,11] A. Kamra *et al.* have theoretically demonstrated this new perspective by accounting the Rayleigh dissipation function in a two-sublattice magnetic system, and the resultant Gilbert damping parameter is independent of temperature near the $T_A$.[11] This damping parameter denoted by $\alpha_{FiM}$ is defined as follows:

$$\alpha_{FiM} = \left| \frac{s_{net}}{s_{total}} \right| \alpha_{FM}, \quad (1)$$

where the $s_{net}$ and $s_{total}$ are the net and total angular momentum, respectively. The $s_{net}$ is calculated by the difference of the angular momentum between two sublattices ($s_{net} = |s_1 - s_2|$) and $s_{total}$ is calculated by the total magnitude of the angular momentum ($s_{total} = |s_1| + |s_2|$). D.-H. Kim *et al.* have experimentally studied the current-driven domain wall motion in ferrimagnetic metal alloy GdFeCo and revealed that the damping parameter $\alpha_{FiM}$ is indeed independent of temperature near the $T_A$.[10] Furthermore, T. Okuno *et al.* has reported that $\alpha_{FiM}$ of the GdFeCo is temperature independent when the FMR measurement temperature is approaching the $T_A$.[6] The FMR of ferrimagnetic thin films below the $T_A$ is difficult to achieve because of much enhanced perpendicular magnetic anisotropy at lower temperatures. It would be desirable that a full temperature range of FMR can be investigated for ferrimagnets.

The divergence of the conventionally defined damping parameter $\alpha_{FM}$ at $T_A$ can be understood easily by considering the energy dissipation rate given by $P = \alpha_{FM} s_{net} \dot{\boldsymbol{m}}^2$ (which is twice the Rayleigh dissipation function), where $\boldsymbol{m}$ is the unit magnetization vector. For the given power $P$ that is pumped into the ferrimagnet by e.g., applying microwave for FMR, as the temperature approaches $T_A$, the net spin density $s_{net}$ decreases and thus $\alpha_{FM}$ increases. Exactly at $T_A$, the net spin density vanishes, making $\alpha_{FM}$ diverge and thus ill-defined. Note that the divergence of $\alpha_{FM}$ at $T_A$ is due to the appearance of the net spin density $s_{net}$ in the dissipation rate and should not be interpreted to indicate the divergence of the dissipation rate, which is always finite. In terms of the alternative damping parameter $\alpha_{FiM}$, the energy dissipation rate is given by $P = \alpha_{FiM} s_{tot} \dot{\boldsymbol{m}}^2$. The total spin density $s_{tot}$ is always finite and has weak temperature dependence, and thus $\alpha_{FiM}$ is well-defined at all temperatures with possibly weak temperature dependence. This suggests that $\alpha_{FiM}$, which is well-defined at all temperatures, might be more useful to describe the damping of ferrimagnetic dynamics, particularly in the vicinity of $T_A$, than the more conventional $\alpha_{FM}$ which diverges and thus ill-defined at $T_A$. One way to appreciate the physical meaning of $\alpha_{FiM}$ is to consider a special model, where the energy dissipation of a ferrimagnet occurs independently through the dynamics of each sublattice and all the sublattices have the same damping parameter. In this case, $\alpha_{FiM}$ is nothing but the damping parameter of the sublattices. So far, the discussion of ferrimagnetic damping is limited to ferrimagnetic metals, while ferrimagnetic insulators have shown the potential for ultralow-power spintronics.[12,13,14,15,16]

In this paper, we investigate the temperature dependence of damping parameters in ferrimagnetic insulator, gadolinium iron garnet ($Gd_3Fe_5O_{12}$, GdIG), by surveying the literature of studies on the temperature dependence of FMR. Since the $s_{total}$ is usually not given in the literature, we adopt two different methods to calculate the individual sublattice magnetization ($M_{Fe}$ and $M_{Gd}$) and then evaluate $s_{total}$. The first method is to use the magnetization of yttrium iron garnet ($Y_3Fe_5O_{12}$, YIG) as the $M_{Fe}$ as done in Ref.[17], where nuclear magnetic resonance experiments show that the magnetization contribution from iron is similar in YIG and GdIG since yttrium does not contribute the magnetization in YIG, and then obtain $M_{Gd}$ from the net magnetization and $M_{Fe}$. The second method uses Brillouin-like function to simulate the temperature dependence of GdIG magnetization, the angular momentum of each individual sublattice can be calculated with the Brillouin function. We found consistent results between these two different methods that the damping parameter $\alpha_{FiM}$ is almost temperature-independent near the $T_A$, unlike the conventionally defined $\alpha_{FM}$ which is strongly temperature-dependent and diverge at $T_A$.

The FMR linewidth ($\Delta H$) of GdIG is utilized to find the conventional damping parameter $\alpha_{FM}$:

$$\Delta H = \frac{\alpha_{FM}}{g_{eff}\mu_B/\hbar} f_{res} + \Delta H_0 , \qquad (2)$$

where $g_{eff}$ is the effective Landé g-factor, $\mu_B$ is the Bohr magneton, $\hbar$ is the reduced Planck constant, $\Delta H_0$ is the frequency-independent inhomogeneous broadening linewidth, and $f_{res}$ is the resonance frequency. Then, to convert the $\alpha_{FM}$ to the $\alpha_{FiM}$, we need to find the ratio $\frac{s_{net}}{s_{total}}$. Note that $\alpha_{FM}$ diverges as the temperature approaches $T_A$, meaning that Eq. (2) can be used only when it is sufficiently far away from the $T_A$. Therefore, we will only employ data sufficiently far away from $T_A$ in this perspective. The net spin density $s_{net}$ is calculated from the difference between the angular momentum of Fe and Gd:

$$s_{Fe} = \frac{M_{Fe}}{g_{Fe}\mu_B/\hbar},$$

$$s_{Gd} = \frac{M_{Gd}}{g_{Gd}\mu_B/\hbar}, \qquad (3)$$

$$s_{net} = |s_{Fe} - s_{Gd}| = \frac{M_{net}}{g_{eff}\mu_B/\hbar},$$

where the $M_{net}$ is the net magnetization, $g_{Fe}$ and $g_{Gd}$ is the Landé g-factor of the iron and gadolinium sublattice, respectively. The net magnetization is given by

$$M_{net} = |M_{Fe} - M_{Gd}|, \qquad (4)$$

which is normally measured by a superconducting quantum interference device or a vibrating-sample magnetometer and provided in the literature.[18,19,20]

**METHOD 1**

We can use the magnetization of YIG as an approximation for the $M_{Fe}$ to calculate the $M_{Fe}$ and $M_{Gd}$ from GdIG net magnetization, as yttrium does not contribute to the magnetization of YIG, which we refer to as ***Method 1***. Experimentally, Boyd et al.[17] used the nuclear ferromagnetic resonance technique to determine temperature-dependent $M_{Fe}$ in YIG and GdIG and found that they are very similar. This approximation has been used in previous literature and has produced reasonable results.[21] The magnetization of YIG is obtained from Ref.18. With $M_{Fe}$ and $M_{Gd}$ known, we can determine the angular momentum of each sublattice with its respective g-factor. The g factors of Fe and Gd are very similar, the g-factor of iron in measured from YIG and is determined as $g_{Fe} = 2.0047$.[22] The g-factor of Gd sublattice is $g_{Gd} = 1.994$ and is determined by measurement of GdIG [23]. The $T_M$ and $T_A$ will be very close to each other, with $T_A$ slightly higher than $T_M$. We can calculate the total spin density $s_{total}$ using

$$s_{total} = s_{Fe} + s_{Gd}. \tag{5}$$

The net spin density $s_{net}$ can be calculated using Eq. (3) and we can obtain effective g-factor meanwhile. Finally, we can calculate the $\alpha_{FM}$ using Eq. (2) and the $\alpha_{FiM}$ using Eq. (1).

**METHOD 2**

The second method is to use the Brillouin-like function to simulate the temperature dependence of magnetization.[24] Due to the weak coupling of the Gd-Gd interaction, the gadolinium magnetic moments follow a paramagnetic behavior and increase drastically at low temperatures. The net magnetization in GdIG can be described by the sum of the three sublattices with a and d sublattices corresponding to Fe and c sublattice corresponding to Gd:

$$M_{net} = |M_a + M_c - M_d|. \tag{6}$$

The individual magnetization component can be simulated by the Brillouin function $B_{S_i}(x_i)$

$$M_i(T) = M_i(0)B_{S_i}(x_i). \tag{7}$$

The $M_i(0)$ is the individual magnetization at 0 K.

$$\begin{aligned} M_d(0) &= 3nm_{Fe} = 3ng_d S_d \mu_B, \\ M_a(0) &= 2nm_{Fe} = 2ng_a S_a \mu_B, \\ M_c(0) &= 3nm_{Gd} = 3ng_c S_c \mu_B, \end{aligned} \tag{8}$$

$n$ is the number of GdIG formula unit per unit volume, it can be calculated using $N_A/(\rho M_r)$, where $N_A$ is the Avogadro's number, $\rho$ and $M_r$ are the density (6.45 $gcm^{-3}$ [25]) and molar mass (942.97) of GdIG respectively. $S_i$ is the electron spin of the respective sublattice. For GdIG, $S_d$ and $S_a$ are 5/2 and $S_c$ is 7/2. $g_i$ is the individual g factor and $x_i$ is defined as:

$$x_d = \left(\frac{\mu_0 S_d g_d \mu_B}{k_B T}\right)(n_{dd}M_d + n_{da}M_a + n_{dc}M_c),$$

$$x_a = \left(\frac{\mu_0 S_a g_a \mu_B}{k_B T}\right)(n_{ad}M_d + n_{aa}M_a + n_{ac}M_c), \qquad (9)$$

$$x_c = \left(\frac{\mu_0 S_c g_c \mu_B}{k_B T}\right)(n_{cd}M_d + n_{ca}M_a + n_{cc}M_c),$$

$n_{ij}$ are the Weiss coefficients between two sublattices, which account for the intersublattice molecular field coupling $(i \neq j)$ or intrasublattice molecular field interactions $(i = j)$.[24] $\mu_0$ is permeability of vacuum.

To determine the $s_{net}$ and $s_{total}$ from the magnetization fitting will require the sublattice g-factor $g_{Gd}$ and $g_{Fe}$. $g_{Fe}$ in a and d sublattice can be experimentally measured from YIG and is determined as $g_{Fe,d} = 2.0047, g_{Fe,a} = 2.003$.[22] The g-factor of Gd c sublattice has the same value as the one in Method 1, $g_{Gd} = 1.994$.[23] With the value of the individual sublattice g-factor, the angular momentum of each sublattice can be calculated from Eq. (3). Then we can calculate the effective gyromagnetic ratio and effective g-factor with the sublattice magnetization and angular momentum.

$$\gamma_{eff} = \frac{M_{Fe,d} - M_{Fe,a} - M_{Gd,c}}{s_{Fe,d} - s_{Fe,a} - s_{Gd,c}}, \qquad (11)$$

$$g_{eff} = \frac{\gamma_{eff} \hbar}{\mu_B},$$

The ratio $s_{net}/s_{total}$ can be calculated where $s_{net} = |s_{Fe,d} - s_{Fe,a} - s_{Gd,c}|$ and $s_{total} = s_{Fe,d} + s_{Fe,a} + s_{Gd,c}$, with both the $g_{eff}$ and angular momentum known. Eventually, the value of $\alpha_{FiM}$ is obtained from Eq. (1).

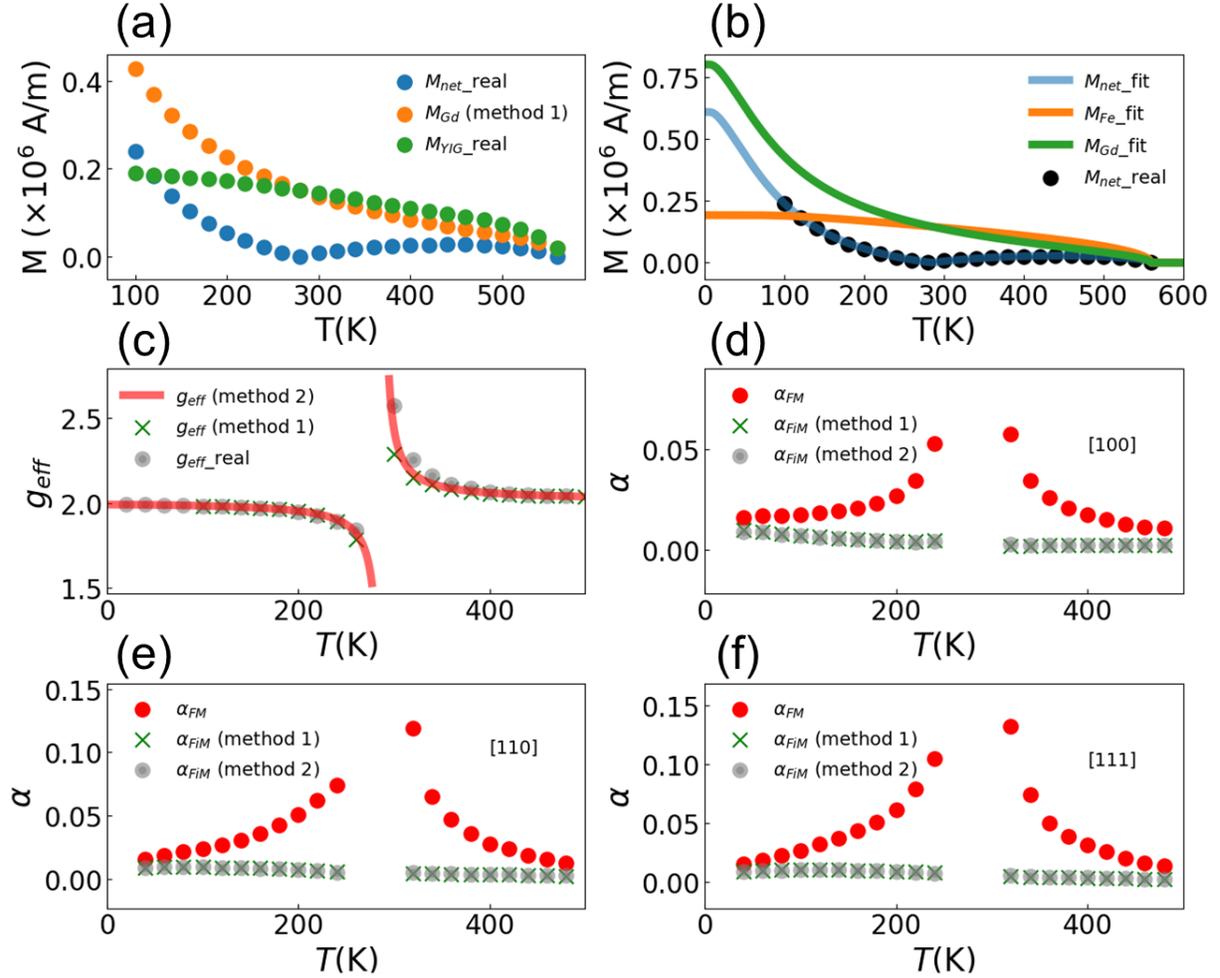

**Figure 1**. The analysis of GdIG data from Rodrigue et al.[23] and Dionne et al.[18] (a) Calculated individual magnetization as a function of temperature using Method 1. (b) The Magnetization curve of GdIG using Brillouin fitting method (Method 2) compared to the magnetization from Dionne et al.[18] (c) The $g_{eff}$ of GdIG calculated from Method 1 as the green cross and from Method 2 as the red line compared to the grey dot $g_{eff}$ from Rodrigue et al. ([100] direction).[23] (d) (e) (f) Comparing the damping parameter $\alpha_{FM}$ (red dot) to $\alpha_{FiM}$ based on Method 1 (green cross) and $\alpha_{FiM}$ based on Method 2 (grey dot) for three directions ([100], [110] and [111]).

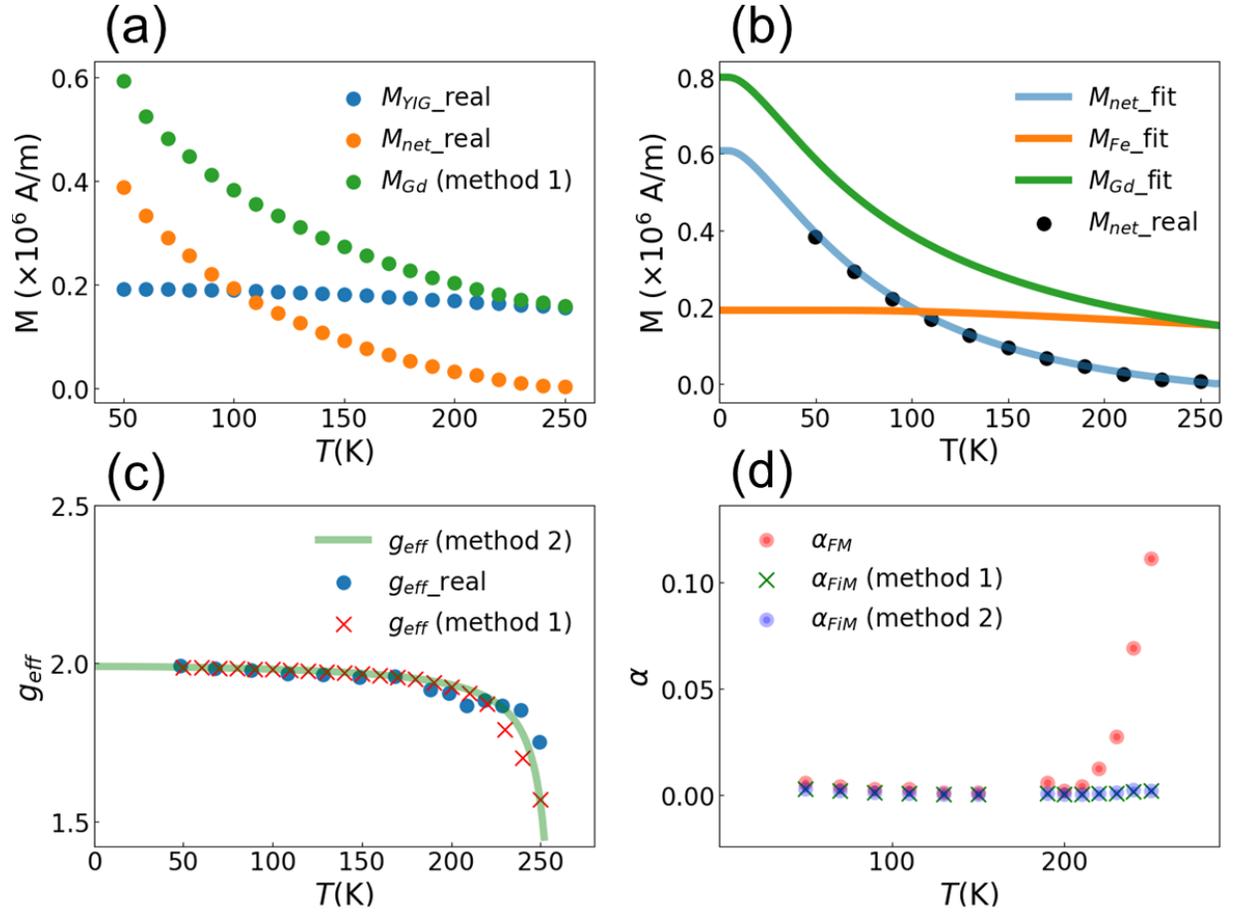

**Figure 2**. The analysis of GdIG data from Flaig et al.[26] (a) Calculated individual magnetization as a function of temperature using Method 1. (b) The Magnetization curve of GdIG using Brillouin fitting method (Method 2) compared to the magnetization from Flaig et al.[26] (c) The $g_{eff}$ of GdIG calculated from Method 1 as the red cross and from Method 2 as the green line compared to the blue dot $g_{eff}$ from Flaig et al.[26] (d) Comparing the damping parameter $\alpha_{FM}$ (red dot) to $\alpha_{FiM}$ based on Method 1 (green cross) and $\alpha_{FiM}$ based on Method 2 (blue dot).

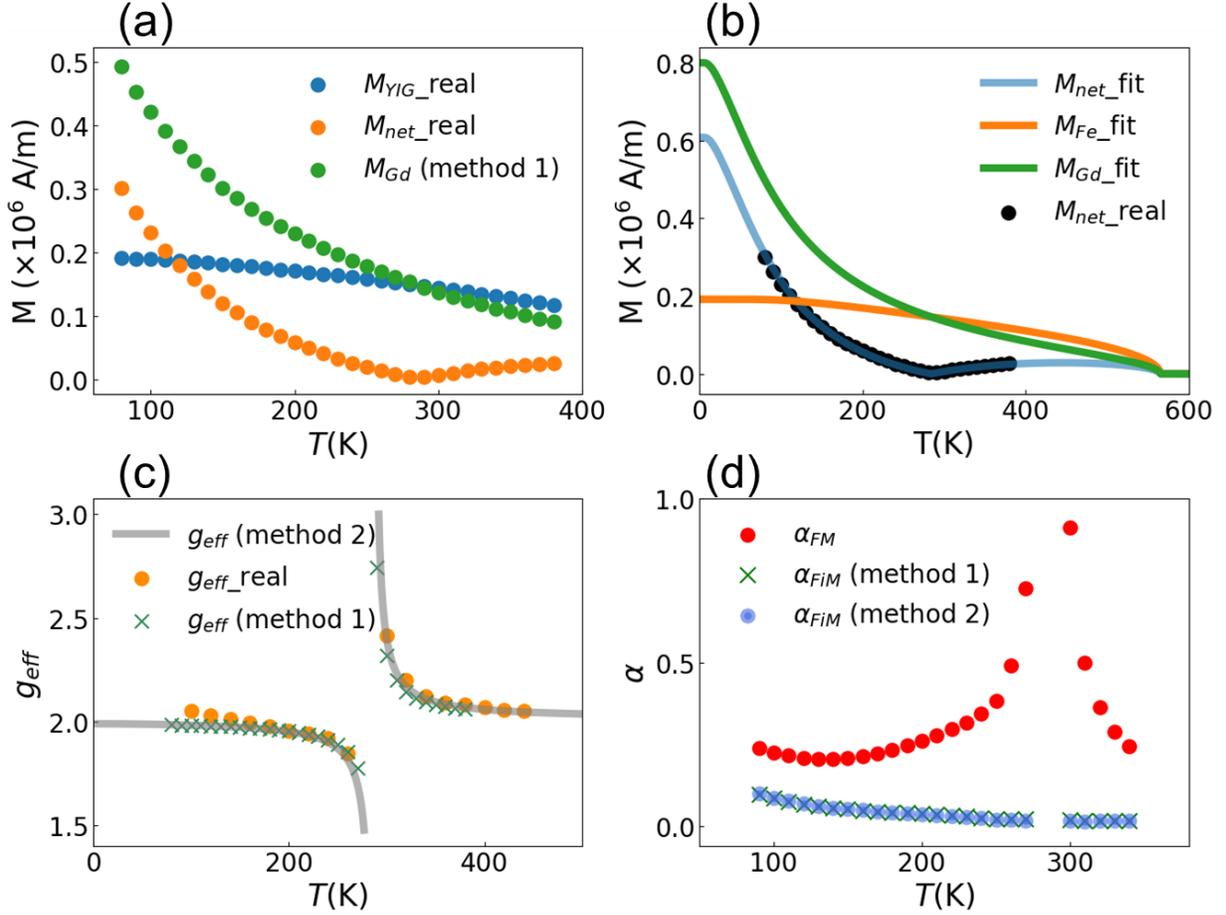

**Figure 3**. The analysis of GdIG data from Calhoun et al[19,20]. (a) Calculated individual magnetization as a function of temperature using Method 1. (b) The Magnetization curve of GdIG using Brillouin fitting method (Method 2) compared to the magnetization from Calhoun *et al*.[20] (c) The $g_{eff}$ of GdIG calculated from Method 1 as the green cross and from Method 2 as the grey line compared to the yellow dot $g_{eff}$ from Calhoun *et al*.[19] (d) Comparing the damping parameter $\alpha_{FM}$ (red dot) to $\alpha_{FiM}$ based on Method 1 (green cross and $\alpha_{FiM}$ based on Method 2 (blue dot).

**RESULTS AND DISCUSSIONS**

We analyze three datasets and evaluate the validity of Method 1 and Method 2 using the formula provided above. The first dataset is from Rodrigue *et al*.[23], where the ΔH and $g_{eff}$ in three directions [100], [110], [111] are provided. Note that the value of $g_{eff}$ is calculated using the Kittel equation in Rodrigue's paper. The $M_{net}$ is obtained from Dionne *et al*.[18] where the GdIG has a similar compensation temperature to Rodrigue *et al*.[23]. $f_{res}$=9.165GHz and we assume that $\Delta H_0$ is zero since the GdIG is a polished sphere. We analyze the data using Method 1 and Method 2 and plot the results in Fig. 1. For Method 1, we can observe that the calculated temperature dependence of the $M_{Gd}$ (see Fig. 1a) and the obtained g-factor (see Fig. 1c) are reasonable. Using Method 2, we get the fitting curves for magnetization from each sublattice and g-factor, which fit accurately to the experimental data.

The second dataset of the temperature dependence of FMR below the $T_A$ is from Maier-Flaig *et al.*,[26] where the g-factor, $\Delta H$, and $M_{net}$ are also provided. Again, we can see that the magnetization as a function of temperature from two methods are in accordance with Flaig's data (see Fig. 2). $g_{eff}$ calculated dots from Method 1 and fitting curves from Method 2 are highly consistent with the data, which illustrates that both two methods are well established.

The third set of data is from B. A. Calhoun *et al.*,[19,20] where $f_{res}$= 9.479GHz. Similar results to the above two datasets are obtained as shown in Fig. 3.

To directly compare the above two methods, the ferrimagnetic damping parameter $\alpha_{FiM}$ calculated from these two methods are plotted against each other in Fig. 1, 2 and 3, using the data from Rodrigue *et al.*[23], Flaig *et al.*[26] and Calhoun *et al.*[19]. For all datasets, two different methods all give consistent results and have similar values: the newly defined damping parameter $\alpha_{FiM}$ of a ferrimagnetic material is not divergent near the $T_A$ and has much lower value than $\alpha_{FM}$. The $\alpha_{FiM}$ in all three datasets is at low value, revealing the achievability of fast domain-wall dynamics in ferrimagnetic insulator at the angular momentum compensation temperature.

**CONCLUSION**

In this work, we survey the literature dataset of FMR studies on the ferrimagnetic insulator GdIG and find that the ferrimagnetic damping parameter $\alpha_{FiM}$ does not increase when the temperature approaches the $T_A$, differing from the conventionally defined $\alpha_{FM}$ that shows divergence near the $T_A$. This validates the recently developed theory about damping in the ferrimagnetic systems and reveals that the damping parameter, when it is appropriately defined with no divergence at all temperatures, is not as high as previously thought. Our work suggests that analyzing the dynamics of ferrimagnets needs extra caution, that is not required for ferromagnets, in particular in the vicinity of the $T_A$ to avoid unphysical divergences. Besides, potentially lower damping in insulators suggests that ferrimagnetic insulators are promising for future ultrafast and ultralow-power spintronic applications.

**ACKNOWLEDGEMENT**

The authors at HKUST were supported by the Hong Kong Research Grants Council-Early Career Scheme (Grant No. 26200520) and the Research Fund of Guangdong-Hong Kong-Macao Joint Laboratory for Intelligent Micro-Nano Optoelectronic Technology (Grant No. 2020B1212030010). S.K.K. was supported by Brain Pool Plus Program through the National Research Foundation of Korea funded by the Ministry of Science and ICT (Grant No. NRF-2020H1D3A2A03099291) and by the National Research Foundation of Korea funded by the Korea Government via the SRC Center for Quantum Coherence in Condensed Matter (Grant No. NRF-2016R1A5A1008184).